\documentclass{amsart}

\usepackage{amsmath}
\usepackage{amssymb}
\usepackage{algorithm}
\usepackage{algorithmic}
\usepackage{upgreek}
\usepackage[hyphens]{url}
\usepackage{hyperref}
\usepackage{breakurl}

\newcommand{\us}{$\upmu$s}
\newcommand{\M}{M}
\newcommand{\MM}{M\!M}

\numberwithin{equation}{section}

\begin{document}

\title{A fast algorithm for reversion of power series}

\author{Fredrik Johansson}
\address{Research Institute for Symbolic Computation \\ Johannes Kepler University \\ 4040 Linz, Austria}
\curraddr{}
\email{fredrik.johansson@risc.jku.at}
\thanks{Supported by Austrian Science Fund (FWF) grant Y464-N18.}

\subjclass[2010]{Primary 68W30}

\date{}

\begin{abstract}
We give an algorithm for reversion of formal power series,
based on an efficient way to implement the Lagrange inversion formula.
Our algorithm requires $O(n^{1/2}(\M(n) + \MM(n^{1/2})))$
operations where $\M(n)$ and $\MM(n)$ are the costs of polynomial and matrix multiplication respectively. This matches the asymptotic complexity of an algorithm of Brent and Kung, but we achieve a constant factor speedup whose magnitude depends on the polynomial and matrix multiplication algorithms used. Benchmarks confirm that the algorithm performs well in practice.
\end{abstract}

\maketitle

\section{Introduction}

Classical algorithms for composition and reversion of power series truncated to order $n$ require $O(n^3)$ operations on elements in the coefficient ring \cite{Knuth1998vol2}. This can be improved to $O(n M(n))$ where $\M(n)$ is the cost of multiplying polynomials of degree less than $n$. In \cite{BrentKung1978}, Brent and Kung gave two asymptotically faster algorithms for composition, and observed that any algorithm for composition can be used for reversion (and vice versa) via Newton iteration, with at most a constant factor slowdown.

The first algorithm (BK 2.1) requires $O(n^{1/2} (\M(n) + \MM(n^{1/2})))$ operations where $\MM(n)$ is the complexity of multiplying two $n \times n$ matrices. This reduces to $O(n^{1/2} \M(n) + n^2)$ with classical matrix multiplication, and $O(n^{1/2} \M(n) + n^{1.91})$ with the Strassen algorithm. The last term can be improved to $O(n^{1.688})$ with the Coppersmith-Winograd algorithm \cite{vonzurGathenGerhard2003}, or $O(n^{1.68632})$ with the recent bound for $\MM(n)$ by Stothers and Vassilevska Williams \cite{Stothers2010, VassilevskaWilliams2011}. The best available bound for the last term is $O(n^{1.667})$, obtained by Huang and Pan \cite{HuangPan1998} using improved techniques for multiplication of nonsquare matrices.

The second algorithm (BK 2.2) requires $O((n \log n)^{1/2} \M(n))$ operations. This is asymptotically slower than BK 2.1 when classical ($\M(n) = O(n^2)$) or Karatsuba multiplication $(\M(n) = O(n^{\log_2{3}}) = O(n^{1.585}))$ is used, but faster when FFT polynomial multiplication $(\M(n) = O(n \log^{1+o(1)} n))$ is available \cite{Bernstein2008, vonzurGathenGerhard2003}.

As noted by Brent and Kung, many special left-compositions, including the evaluation of reciprocals, square roots, and elementary transcendental functions of power series, can be done in merely $O(\M(n))$ operations. Recent research has focused on speeding up such evaluations by constant factors by eliminating redundancy from Newton iteration \cite{Bernstein2004, HanrotZimmermann2004, Harvey2011}. Algorithms with quasi-linear complexity are also known for certain right-compositions, including right-composition by algebraic functions \cite{vdH:relax} and some special functions \cite{BostanSalvySchost2008}. Improved composition algorithms over special rings have also been investigated \cite{Bernstein1998, KedlayaUmans2011, Ritzmann1986}. However, the algorithms of Brent and Kung have remained the best known for composition and reversion in the general case.

In this paper, we give a new algorithm for reversion analogous to BK~2.1 and likewise requiring $O(n^{1/2} (\M(n) + \MM(n^{1/2})))$ operations, but achieving a constant factor speedup. The speedup ratio depends on the asymptotics of $M(n)$ and $\MM(n)$ and is in the range between 1.2 and 2.6 with polynomial and matrix multiplication algorithms used in practice. Our algorithm also allows incorporating the complexity refinement of Huang and Pan.

Whereas BK~2.1 can be viewed as a baby-step giant-step version of Horner's rule, our algorithm can be viewed as a baby-step giant-step version of the Lagrange inversion formula, avoiding Newton iteration entirely (apart from a single $O(M(n))$ reciprocal computation). It is somewhat surprising that such an algorithm has been overlooked until now, with all publications following Brent and Kung apparently having taken Newton iteration as the final word on the subject matter.

\section{The algorithm}

Our setting is the ring of truncated power series $R[[x]]/\langle x^n\rangle$ over a commutative coefficient ring $R$ in which the integers $1, \ldots, n-1$ are cancellable (i.e. nonzero and not zero divisors). For example, we may take $R = \mathbb{Z}$ or $R = \mathbb{Z}/p\mathbb{Z}$ with prime $p \ge n$. We recall the Lagrange inversion formula (\cite{Knuth1998vol2}, p. 527). If $f(x) = x / h(x)$ where $h(0)$ is a unit in $R$, then the compositional inverse or reversion $f^{-1}(x)$ satisfying $f(f^{-1}(x)) = f^{-1}(f(x)) = x$ exists and its coefficients are given by
\begin{equation*}
[x^k] f^{-1}(x) = \frac{1}{k} [x^{k-1}] h(x)^k.
\end{equation*}
The straightforward way to evaluate $n$ terms of $f^{-1}(x)$ with the Lagrange inversion formula is to compute $h(x)$ (this requires $O(M(n))$ operations with Newton iteration) and then compute the powers $h^2, h^3, \ldots$ successively, for a total of $(n + O(1)) M(n)$ operations.

Our observation is that it is redundant to compute all the powers of $h$
given that we only are interested in a single coefficient from each.
To do better, we choose some $1 \le m < n$ and
precompute $h, h^2, h^3, \ldots, h^m$. For $0 \le k < n$, we can then write $h^k$ as $h^{i+j}$ where $0 \le j < m$ and $i = lm$ for some $0 \le l \le \lceil n/m \rceil$, only requiring $h^m, h^{2m}, h^{3m}, \ldots$ to be computed subsequently. Determining a single coefficient in $h^k = h^i h^j$ can then be done in $O(n)$ operations using the definition of the Cauchy product. Picking $m \approx n^{1/2}$ minimizes the number of polynomial multiplications required.

We give a detailed account of this procedure as Algorithm \ref{fastlag}. We note that most of the polynomial arithmetic is done to length $n-1$ rather than length $n$, as the initial coefficient always is zero.

\begin{algorithm}
\caption{Fast Lagrange inversion}
\label{fastlag}
\begin{algorithmic}
\renewcommand{\algorithmicrequire}{\textbf{Input:}}
\renewcommand{\algorithmicensure}{\textbf{Output:}}
\REQUIRE $f = a_1 x + a_2 x^2 + \ldots + a_{n-1} x^{n-1}$ where $n > 1$ and $a_1$ is a unit in $R$
\ENSURE $g = b_1 x + \ldots + b_{n-1} x^{n-1}$ such that $f(g(x)) = g(f(x)) = x \bmod x^n$
\STATE $m \gets \lceil \sqrt{n-1} \rceil$
\STATE $h \gets x / f \bmod x^{n-1}$
\FOR{$1 \le i < m$}
  \STATE $h^{i+1} \gets h^i \times h \bmod x^{n-1}$
  \STATE $b_{i} \gets \frac{1}{i} [x^{i-1}] h^i$
\ENDFOR
\STATE $t \gets h^m$
\FOR{$i = m, 2m, 3m, \ldots, lm < n$}
  \STATE $b_i \gets \frac{1}{i} [x^{i-1}] t$
  \FOR {$1 \le j < m$ while $i + j < n$}
    \STATE $b_{i+j} \gets \frac{1}{i+j} \sum_{k=0}^{i+j-1} ([x^k] t) \cdot ([x^{i+j-k-1}] h^j)$
  \ENDFOR
  \STATE $t \gets t \times h^m \bmod x^{n-1}$
\ENDFOR
\RETURN $b_1 + b_2 x + \ldots + b_{n-1} x^{n-1}$
\end{algorithmic}
\end{algorithm}

\subsection*{An improved version}

Algorithm \ref{fastlag} clearly requires $O(n^{1/2} \M(n) + n^2)$ operations in $R$, as many as BK~2.1 with classical matrix multiplication. We can improve the complexity by packing the inner loops into a single matrix product as shown in Algorithm~\ref{fastlag2}. This allows us to exploit fast matrix multiplication.

\begin{algorithm}
\caption{Fast Lagrange inversion, matrix version}
\label{fastlag2}
\begin{algorithmic}
\renewcommand{\algorithmicrequire}{\textbf{Input:}}
\renewcommand{\algorithmicensure}{\textbf{Output:}}
\REQUIRE $f = a_1 x + a_2 x^2 + \ldots + a_{n-1} x^{n-1}$ where $n > 1$ and $a_1$ is a unit in $R$
\ENSURE $g = b_1 x + \ldots + b_{n-1} x^{n-1}$ such that $f(g(x)) = g(f(x)) = x \bmod x^n$
\STATE $m \gets \lceil \sqrt{n-1} \rceil$
\STATE $h \gets x / f \bmod x^{n-1}$
\STATE Assemble $m \times m^2$ matrices $B$ and $A$ from
$h, h^2, \ldots, h^m$ and $h^m, h^{2m}, h^{3m}, \ldots$.
\FOR{$1 \le i \le m, \;1 \le j \le m^2$}
  \STATE $B_{i,j} \gets [x^{i+j-m-1}] \; h^i$
  \STATE $A_{i,j} \gets [x^{im-j}] \; h^{(i-1)m}$
\ENDFOR
\STATE $C \gets AB^T$
\FOR{$1 \le i < n$}
  \STATE $b_i \gets C_i / i$ ($C_i$ is the $i$th entry of $C$ read rowwise)
\ENDFOR
\RETURN $b_1 + b_2 x + \ldots + b_{n-1} x^{n-1}$
\end{algorithmic}
\end{algorithm}

In the description of Algorithm \ref{fastlag2}, the matrices are indexed from 1 and the pseudocode has been simplified by letting the exponents run out of bounds, using the convention that $[x^k] p = 0$ when $k < 0$ or $k \ge n - 1$. To see that the algorithm is correct, write $[x^{i_1+(i_2-1)m-1}] h^{i_1+(i_2-1)m}$ as
\begin{equation*}
\sum_{j=0}^{i_1+(i_2-1)m-1} \left(\left[ x^j \right] h^{i_1}\right) \left(\left[x^{i_1 + (i_2-1)m-1-j}\right] h^{(i_2-1)m}\right)
\end{equation*}
and shift the summation index to obtain
\begin{equation*}
\sum_{j=m-i_1+1}^{i_2 m} \left(\left[x^{i_1+j-m-1}\right] h^{i_1}\right)
\left(\left[x^{i_2 m-j}\right] h^{(i_2-1)m}\right)
\end{equation*}
which is the inner product of the nonzero part of row $i_1$ in $B$ with the nonzero part of row $i_2$ in $A$.

The structure of the matrices is perhaps illustrated more clearly by an example. Taking $n = 8$ and $m = 3$, we need the coefficients of $1, x,\ldots,x^6$ in powers of $h$. Letting $h^k_i$ denote $[x^i] h^k$, the matrices become
\begin{equation*}
A =
\begin{pmatrix}
h^0_2 & h^0_1 & h^0_0 & 0 & 0 & 0 & 0 & 0 & 0 \\
h^3_5 & h^3_4 & h^3_3 & h^3_2 & h^3_1 & h^3_0 & 0 & 0 & 0 \\
(h^6_8) & (h^6_7) & h^6_6 & h^6_5 & h^6_4 & h^6_3 & h^6_2 & h^6_1 & h^6_0 
\end{pmatrix}
\end{equation*}

\begin{equation*}
B =
\begin{pmatrix}
0 & 0 & h^1_0 & h^1_1 & h^1_2 & h^1_3 & h^1_4 & h^1_5 & h^1_6 \\
0 & h^2_0 & h^2_1 & h^2_2 & h^2_3 & h^2_4 & h^2_5 & h^2_6 & (h^2_7) \\
h^3_0 & h^3_1 & h^3_2 & h^3_3 & h^3_4 & h^3_5 & h^3_6 & (h^3_7) & (h^3_8)
\end{pmatrix}
\end{equation*}
where entries in parentheses do not contribute to the final result and may be set to zero. In this example the coefficient of $x^4$ in $h^5$ is given by the fifth entry in $C$, namely $C_{2,2} = h^3_4 h^2_0 + h^3_3 h^2_1 + h^3_2 h^2_2 + h^3_1 h^2_3 + h^3_0 h^2_4$.

\section{Complexity analysis}

We now study the complexity in some more detail. Let $m = \lceil \sqrt{n-1} \rceil$. Then Algorithm \ref{fastlag2} involves at most:

\begin{enumerate}
\item $2m+O(1)$ polynomial multiplications, each with cost $M(n)$
\item One $(m \times m^2)$ times $(m^2 \times m)$ matrix multiplication
\item $O(n)$ additional operations
\end{enumerate}

For comparison, BK 2.1 requires at most:

\begin{enumerate}
\item $m$ polynomial multiplications, each with cost $M(n)$
\item One $(m \times m)$ times $(m \times m^2)$ matrix multiplication
\item $m$ polynomial multiplications and additions, each with cost $M(n) + n$
\end{enumerate}

Brent and Kung break the matrix multiplication into $m$ products of $m \times m$ matrices, requiring $m \MM(m)$ operations. We can do the same in Algorithm~\ref{fastlag2}, writing the product as a length-$m$ inner product of $m \times m$ matrices. The extra $O(n^{3/2})$ additions in this matrix operation do not affect the complexity, but it is interesting to note that they match the $O(n^{3/2})$ additions in the last polynomial stage of BK~2.1. To summarize, both Algorithm \ref{fastlag2} and BK~2.1 require at most $(2n^{1/2}+O(1)) M(n) + n^{1/2}\MM(n^{1/2}) + O(n^{3/2})$ operations.

The primary drawback of our algorithm as well as BK 2.1 is the requirement to store $O(n^{3/2})$ temporary coefficients in memory, compared to $O(n \log n)$ for BK 2.2 and $O(n)$ for a naive implementation of Lagrange inversion.

\subsection*{Avoiding Newton iteration.}

In effect, we need the same number of operations to perform a length-$n$ reversion with fast Lagrange inversion as to perform a length-$n$ composition with BK~2.1. However, to perform a reversion with BK~2.1, we must employ Newton iteration. Using the update
\begin{equation*}
g_{k+1}(x) = \frac{f(g_k(x)) - x}{f'(g_k(x))},
\end{equation*}
where the chain rule allows us to reuse the composition in the numerator for the denominator, this entails computing a sequence of compositions of length $l = 1, \ldots, \lceil n/4 \rceil, \lceil n/2 \rceil, n$, plus a fixed number of polynomial multiplications of the same length at each stage. If $c$ and $r$ are such that a length-$n$ composition takes $C(n) = cn^{r}$ operations, Newton iteration asymptotically takes
\begin{equation*}
C(n) + C(n/2) + C(n/4) + \ldots = cn^r \left( \frac{2^r}{2^r-1} \right)
\end{equation*}
operations, ignoring additional polynomial multiplications. For example, with classical polynomial multiplication as the dominant cost ($r = 5/2$), the speedup given by the expression in parentheses is $\frac{4}{31}(8+\sqrt{2}) \approx 1.214$ . With FFT polynomial multiplication, and classical matrix multiplication as the dominant cost ($r = 2$), the speedup is $4/3$. We note that a more efficient form of the Newton iteration might exist, in which case the speedup would be smaller.

\subsection*{Improving the matrix multiplication.}

If the matrix multiplication dominates, we can gain an additional speedup from the fact that the $i$th $m \times m$ block of the matrix $A$ only has $m - i + 1$ nonzero rows, whereas the matrices in BK 2.1 are full. Classically this gives a twofold speedup, reflected in the loop boundaries of Algorithm \ref{fastlag}. We should ideally modify Algorithm \ref{fastlag2} to include this saving.

In fact, a speedup is attainable with any square matrix multiplication algorithm $\MM(m) \sim m^{\omega}$ where $\omega > 2$. For simplicity, assume that $m$ is sufficiently composite. Do the first $m/2$ products as full products of size $m$, the next $(m/2 - m/3)$ in blocks of size $m/2$, the next $(m/3 - m/4)$ in blocks of size $m/3$, and so on. At stage $k$, only $k^2$ products of blocks of size $m/k$ are required. The speedup achieved through this procedure is
\begin{equation*}
m^{\omega+1} \left( \sum_{k=1}^{\infty} \left(\frac{m}{k} - \frac{m}{k+1}\right) k^2 \left(\frac{m}{k}\right)^{\omega}\right)^{-1} > \left(\sum_{k=0}^{\infty} \frac{2^{k-1}}{2^{k\omega}}\right)^{-1} = 2 - 2^{2-\omega} > 1
\end{equation*}
where the nontrivial inequality can be obtained by considering the analogous subdivison with blocks of size $m/2^k$ only.

Alternatively, we can write $AB^T = (AP) (P^{-1} B^T)$ where $P$ is a permutation matrix that makes each $m \times m$ block in $A$ lower triangular, and use any algorithm that speeds up multiplication between a full and a triangular matrix. A simple recursive decomposition of size-$k$ blocks into size-$k/2$ blocks has a proportional cost of $C(k) = 4C(k/2) + 2(k/2)^{\omega} + O(k^2)$, providing a speedup of $2^{\omega-1}-2$. This is greater than 1 when $\omega > \log_2 6 \approx 2.585$, and better than the first method when $\omega > 1+\log_2(2+\sqrt{2}) \approx 2.771$. In particular, we recover an optimal factor-two speedup with classical multiplication, and a $3/2$ speedup with the Strassen algorithm.

\subsection*{Using rectangular multiplication}

Let $n = m^2$. In the preceding analysis, we have multiplied $m \times n$ matrices via decomposition into square blocks. Remarkably, Huang and Pan have shown \cite{HuangPan1998} that this is not asymptotically optimal with the best presently known algorithms. Letting $\MM(x,y,z)$ denote the complexity of multiplying a matrix of size $x \times y$ by a matrix of size $y \times z$, Huang and Pan show that $\MM(m,m,n) = O(n^{1.667})$, improving on the best known bound $m \MM(m,m,m) = O(n^{1.68632})$ obtained via multiplication of square matrices.

The complexity improvement of Huang and Pan also translates to Algorithm~\ref{fastlag2}. More precisely, given any algorithm for $m \times m$ by $m \times n$ matrix multiplication over a general ring, there is a transposed version for $m \times n$ by $n \times m$ matrix multiplication that uses the same number of scalar multiplications \cite{Hopcroft1973} and a number of extra scalar additions bounded by the number of entries \cite{Probert1976}. We can therefore take $\MM(m,n,m) = (1+o(1))\MM(m,m,n)$.

With the Huang-Pan algorithm, we do not know whether a constant factor can be saved by exploiting the zero entries. This problem would be interesting to explore further. In any case, the Huang-Pan algorithm is currently only of theoretical interest, as the advantage probably only can be realized for infeasibly large matrices.

\subsection*{Practical performance}

\begin{table}
\begin{center}
\begin{tabular}{|c|c|c|c|c|}
\hline
Dominant operation & Complexity & Newton & Matrix & Total \\
\hline
Polynomial, classical & $O(n^{5/2})$ & 1.214 & 1 & 1.214 \\
Polynomial, Karatsuba & $O(n^{1/2+\log_2 3})$ & 1.308 & 1 & 1.308 \\
Matrix, classical & $O(n^2)$ & 1.333 & 2.000 & 2.666 \\
Matrix, classical, $n$-bit coeff. & $O(n^{3+o(1)})$ & 1.142 & 2.000 & 2.285 \\
Matrix, Strassen & $O(n^{(1+\log_2 7)/2})$ & 1.364 & 1.500 & 2.047 \\
Matrix, Cop.-Win. & $O(n^{1.688})$ & 1.450 & 1.229 & 1.782 \\
Matrix, Huang-Pan & $O(n^{1.667})$ & 1.458 & 1? & 1.458? \\
(Polynomial, FFT)* & $O(n^{3/2+o(1)})$ & 1.546 & 1 & 1.546 \\
(Polynomial, FFT, $n$-bit coeff.)* & $O(n^{5/2+o(1)})$ & 1.214 & 1 & 1.214 \\
\hline
\end{tabular}
\end{center}

\caption{Theoretical speedup of Algorithm 2 over BK 2.1 due
to avoiding Newton iteration and exploiting the matrix structure.
*Assuming that matrix multiplication can be ignored.}
\label{tab:speedup}
\end{table}

Table \ref{tab:speedup} gives a summary of the theoretical speedup gained by Algorithm \ref{fastlag2} over BK~2.1 with various matrix and polynomial multiplication algorithms. With FFT-based polynomial multiplication, BK~2.2 is asymptotically faster than BK~2.1 and hence also than Algorithm \ref{fastlag2}. In practice, however, the overhead of quasilinear polynomial multiplication compared to matrix multiplication is likely to be large. Fast Lagrange inversion can therefore be expected to be faster than not only BK~2.1 but also BK~2.2 even for quite large $n$.

Of course, counting ring operations may not accurately reflect actual speed since operations in most interesting rings take a variable amount of time to execute on a physical computer. One consequence of this fact is that Newton iteration is likely to impose a smaller overhead than predicted, since coefficients generally are smaller in earlier steps than in later ones. Newton iteration can also be expected to perform better than generically when the output as a whole has small coefficients.

Over $\mathbb{Z}$ in particular, arithmetic operations with $b$-bit numbers cost $O(b^{1+o(1)})$, where the complexity measure is the number of bit operations. In power series arising in applications, we often have $b = O(n^{1\pm\varepsilon})$. Two complexity estimates based on this assumption are included in Table~\ref{tab:speedup}. In practice, the speed will be sensitive to the sizes of the coefficients appearing internally in each algorithm, varying with the structure of $f(x)$.

We note that fast Lagrange inversion becomes faster than generically when the coefficients of $x / f(x)$ grow slowly. This is often the case when $f(x)$ is a rational function. The reversion of a rational function of fixed degree can be computed faster by a dedicated algorithm (Newton iteration takes $O(M(n))$ operations, using polynomial evaluation and series division for the composition), but it is desirable for a general-purpose algorithm to be efficient in this common case, and Lagrange inversion of course also works for nonrational functions having this growth property.

\section{Benchmarks}

We have implemented tuned versions of naive Lagrange inversion (``Lagrange''), BK 2.1 with Newton iteration, and Algorithm \ref{fastlag} (``Fast Lagrange'') over $\mathbb{Z}/p\mathbb{Z}$, $\mathbb{Z}$ and $\mathbb{Q}$ as part of the FLINT library \cite{Hart2010}. For each of these rings, FLINT provides fast coefficient arithmetic (using MPIR \cite{MPIR} for bignum arithmetic) and asymptotically fast polynomial multiplication using Kronecker substitution and the Sch\"{onhage}-Strassen FFT algorithm. Matrix multiplication over $\mathbb{Z}/p\mathbb{Z}$ uses the Strassen algorithm when the smallest dimension is at least 256, which in principle helps BK~2.1 for $n > 256^2$ (the speedup is not significant for feasible $n$, however).

Timings over $\mathbb{Z}/p\mathbb{Z}$ obtained on an Intel Xeon E5-2650 2.0 GHz CPU with 256 GiB of RAM are given in Table \ref{tab:Zp}. Algorithm~\ref{fastlag} consistently runs about 1.6 times faster than BK 2.1, agreeing with a predicted speedup of 1.546 with quasilinear polynomial multiplication and negligible cost of matrix multiplication -- we see that polynomial multiplication indeed dominates in BK~2.1 for $n$ up to at least~$10^6$. We have also implemented BK~2.2 over $\mathbb{Z}/p\mathbb{Z}$, finding it to take about twice as much time as BK~2.1 in the tested range. BK~2.2 might however be preferable for larger~$n$ due to memory limits (with $n = 10^6$ and 64-bit coefficients, BK~2.1 uses 15 GiB of memory and fast Lagrange reversion uses 7.5 GiB of memory).

\begin{table}[!ht]
\begin{center}
\begin{tabular}{ | l | c | c | c | c| }
\hline
$n$ & Lagrange & BK 2.1 & BK 2.2 & Fast Lagrange \\
\hline
$10$     &   10 \us     &   10   & 18  &  6.1 \\
$10^2$   &   2.8 ms     &  0.92   & 1.6  &  0.54 \\
$10^3$   &   690 ms    &  66   & 120  &  45 \\
$10^4$   &   110 s     &  3.3 (8\%)   & 7.1  &  2.1 \\
$10^5$   &   12100 s   &  144 (20\%)  & 315  &  85 \\
$10^6$   &   $1.9 \cdot 10^6$ s    &  8251 (28\%) & 15131  & 4832 \\
\hline
\end{tabular}
\end{center}
\caption{Timings for reversion of a random power series over $\mathbb{Z}/p\mathbb{Z}, p = 2^{63} + 29$. The percentage of time spent on matrix multiplication in BK~2.1 is shown in parentheses.}
\label{tab:Zp}
\end{table}

Ring operations in $\mathbb{Z}$ and $\mathbb{Q}$ do not take constant time, as reflected in Tables \ref{tab:Z} and~\ref{tab:Q}. Timings are roughly cubic in $n$ as expected from theory, but sensitive to the inputs. Fast Lagrange inversion is the fastest algorithm for small $n$ in all examples, the fastest in all examples over $\mathbb{Z}$, and substantially faster for the rational functions $f_3$ and $f_6$ (in both cases $x / f(x)$ has small coefficients). For large $n$, BK~2.1 performs well on $f_4$ and $f_5$, presumably due to generating smaller coefficients internally.

\begin{table}
\begin{center}
\begin{tabular}{ |l | c c c | c c c | c c c| }
\hline
$n$ & \multicolumn{3}{c|}{Lagrange} &
      \multicolumn{3}{c|}{BK 2.1} & 
      \multicolumn{3}{c|}{Fast Lagrange} \\
        & $f_1$ & $f_2$ & $f_3$
        & $f_1$ & $f_2$ & $f_3$
        & $f_1$ & $f_2$ & $f_3$ \\
\hline
10      & 7.0 \us     & 6.5    & 6.3
        & 10          & 10     & 10
        & 4.9         & 4.3    & 4.1     \\
$10^2$  & 31 ms       & 7.2    & 3.2
        & 7.8         & 2.1    & 2.1
        & 6.4         & 0.96   & 0.65    \\
$10^3$  & 106 s       & 10     & 4.5
        & 10          & 1.1    & 0.96
        & 7.1         & 0.71   & 0.22   \\
        &             &        &
        & (38\%)      & (31\%) & (11\%)
        &             &        &        \\
$10^4$  & -           & -      & -
        & 24356 s     & 1453   & 538
        & 8903        & 426    & 152    \\
        &             &        &
        & (81\%)      & (67\%) & (10\%)
        &             &        &        \\
\hline
\end{tabular}
\end{center}
\caption{Timings for the reversion of ${f_1(x) = \sum_{k\ge1} k! x^k}$,
${f_2(x) = \frac{x}{\sqrt{1-4x}}}$,
${f_3(x) = \frac{x+x^2}{1+x+x^2}}$ over $\mathbb{Z}$.}
\label{tab:Z}
\end{table}
\begin{table}
\begin{center}
\begin{tabular}{ |l | c c c | c c c | c c c| }
\hline
$n$ & \multicolumn{3}{c|}{Lagrange} &
      \multicolumn{3}{c|}{BK 2.1} & 
      \multicolumn{3}{c|}{Fast Lagrange} \\
        & $f_4$ & $f_5$ & $f_6$
        & $f_4$ & $f_5$ & $f_6$
        & $f_4$ & $f_5$ & $f_6$ \\
\hline
10      & 15 \us      & 15      & 13
        & 31          & 28      & 28
        & 11          & 11      & 9.1    \\
$10^2$  & 40  ms      & 40      & 10
        & 12          & 21      & 8.8
        & 8.9         & 8.1     & 1.9   \\
$10^3$  & 145 s       & 133     & 9.7
        & 8.8         & 17      & 3.1
        & 14          & 13      & 0.65   \\
        &             &         &
        & (28\%)      & (24\%)  & (19\%)
        &             &         &        \\
$10^4$  & -           & -       & -
        & 13812 s     & 27057   & 1990
        & 35633       & 27823   & 784    \\
        &             &         &
        & (27\%)      & (27\%)  & (14\%)
        &             &         &        \\
\hline
\end{tabular}
\end{center}
\caption{Timings for the reversion of ${f_4(x) = \exp(x) - 1}$, ${f_5(x) = x \exp(x)}$,
${f_6(x) = \frac{3x(1-x^2)}{2(1-x+x^2)^2}}$ over $\mathbb{Q}$.}
\label{tab:Q}
\end{table}
With larger coefficients (as seen especially in the case of $f_1$), matrix multiplication appears to take a larger proportion of the time, suggesting that BK~2.2 becomes competitive for smaller $n$. We have not implemented BK~2.2 over $\mathbb{Z}$ and~$\mathbb{Q}$, however, and can therefore not provide a direct comparison. 

Care must be taken to handle denominators efficiently. In our implementation of BK~2.1, we found that naive matrix multiplication over $\mathbb{Q}$ took ten times as long as polynomial multiplications. Clearing denominators and multiplying matrices over $\mathbb{Z}$ resulted in a comparable time being spent on the matrix and polynomial stages. Similar concerns apply when implementing Algorithms \ref{fastlag} and \ref{fastlag2}. On the other hand, translating the \emph{entire} composition or reversion to one over $\mathbb{Z}$ by rescaling the inputs typically results in too much coefficient inflation, and can even run slower than a classical algorithm working directly over $\mathbb{Q}$. We expect the situation to be similar when working with parametric power series having rational functions as coefficients.

An interesting alternative would be to work modulo small primes and reconstruct the output using the Chinese remainder theorem. We have not investigated this option in detail. It would provide additional memory benefits: for example, if the coefficients have $O(n)$ bits, direct application of BK~2.1 or fast Lagrange reversion uses $O(n^{5/2})$ bits of temporary space, while modular reversions each require $O(n^{3/2+o(1)})$ bits of space -- less than the $O(n^2)$ bits required to store the output.

\section{Conclusion}

Fast Lagrange inversion is a practical algorithm for reversion of formal power series,
having essentially no higher overhead than a naive implementation of Lagrange
inversion for small $n$ and requiring fewer coefficient operations than Newton
iteration coupled with BK 2.1 for large $n$. Among currently available general-purpose algorithms, it is likely to be the fastest choice for
typical coefficient rings, input series, and values of $n$, and may thus be a good
choice as a default reversion algorithm in a computer algebra system.
Newton iteration with BK 2.2 remains faster asymptotically
when FFT polynomial multiplication is available, and uses less memory, but may require
very large $n$ to become advantageous.

An interesting question is whether a reversion analog of BK~2.2 can be constructed
that avoids Newton iteration, or whether we can otherwise save constant factors in BK~2.2.
We may also ask whether the close correspondence in complexity between Algorithm~\ref{fastlag2} and BK~2.1 can be explained by some underlying duality along the lines of the transposition principle.
Further investigation of improvements over particular rings and of the special matrix multiplications arising in Algorithm \ref{fastlag2} and BK~2.1 would also be warranted.

\section{Acknowledgements}

We thank the anonymous referee for pointing us to previous literature and providing several clarifications and corrections. We also acknowledge the valuable feedback from William Hart and Manuel Kauers.

\bibliographystyle{amsplain}
\bibliography{refs.bib}

\end{document}